\documentclass[aps,prl,amsmath,twocolumn,groupedaddress,showpacs,letterpaper]{revtex4-1}
\usepackage{graphicx}
\usepackage{epstopdf}

\begin{document}
\title{Generation of nonclassical microwave states using an artificial atom in 1D open space}
\author{Io-Chun Hoi, Tauno Palomaki, G\"oran Johansson, Joel Lindkvist, Per Delsing \& C.M.Wilson}

\email{chris.wilson@chalmers.se}

\affiliation{Department of Microtechnology and Nanoscience (MC2), Chalmers University of Technology, G\"oteborg, Sweden.}

\date{\today}

\begin{abstract}

We have embedded an artificial atom, a superconducting transmon
qubit, in a 1D open space and investigated the scattering properties
of an incident microwave coherent state. By studying the statistics of
the reflected and transmitted fields, we demonstrate that the
scattered states can be nonclassical.  In particular, by measuring the second-order correlation
function, $g^{(2)}$, we show photon antibunching
in the reflected field and superbunching in the transmitted field. We also compare the elastically and
inelastically scattered fields using both phase-sensitive and phase-insensitive measurements.

\end{abstract}
 \maketitle
 
 A single atom interacting with propagating electromagnetic fields in open
space is a fundamental system of quantum optics. Strong coupling between a
single artificial atom and resonant propagating fields has recently been achieved
in a 1D system \cite{Astafiev1,Hoi}, experimentally demonstrating nearly perfect extinction of the forward propagating fields \cite{Hoi}. However, this extinction can
be explained by classical theory: a classical point-like oscillating
dipole perfectly reflects resonant incident fields \cite{Zumofen}. In this Letter, we
demonstrate the quantum nature of the transmitted and reflected fields
generated from our artificial atom in 1D open space by using a resonant coherent
state as the incident field. In particular, by measuring the statistics of the fields we show that the reflected field is antibunched \cite{Chang,Loudon} while still
maintaining first-order coherence. Moreover, we observe superbunching
statistics in the transmitted fields \cite{Chang}.

To understand how our artificial atom generates antibunched and superbunched
states, it is helpful to consider the incident coherent state in the photon
number basis. For a low power incident field with less than 0.5 average
photons per lifetime of our atom, we can safely approximate the coherent
field using only the first three photon eigenstates. If we consider a
one-photon incident state, the atom reflects it, leading to
antibunching statistics in the reflected field. Together with the
zero-photon state the reflected field still maintains
first-order coherence. For a two-photon incident state, since the atom is not able to scatter more than one photon at a time, the pair has a much higher probability of transmission, leading to superbunching statistics in transmission \cite{Zheng, Chang}. In this sense, our single artificial atom acts as a photon-number filter, which extracts the one-photon number state from a coherent state.  This represents a novel way to generate photon correlations and nonclassical states in the microwave compared to other recent work \cite{chris,Eichler,Eichler2,Reulet,Mallet}.

Our system consists of a superconducting ÓtransmonÓ
qubit \cite{Koch}, strongly coupled to a 1D transmission line in a
coplanar waveguide configuration(see Fig. 1A). The ground state
$\left\vert 0 \right\rangle$ and first excited state $\left\vert 1
\right\rangle$ have a transition energy
$\hbar\omega_{01}$. The relaxation rate of the atom is dominated by an intentionally
strong coupling to the Z = 50 $\Omega$ transmission line through the
coupling capacitor $C_c$, defined in Fig. 1B.

\begin{figure}
 \includegraphics[width=\columnwidth]{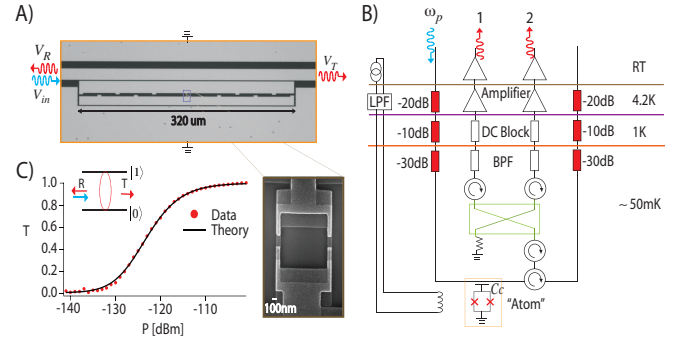}
\caption{(A) A  micrograph of our artificial atom, a superconducting
transmon qubit embedded in a 1D open transmission line. (Zoom In)
Scanning-electron micrograph of the SQUID loop of the transmon.
(B) Schematic setup for measurement of the second-order correlation
function. This setup enables us to do Hanbury
Brown-Twiss measurements between
output ports 1 and 2. Depending on the choice of input port, we
can measure $g^{(2)}$ of the reflected/transmitted field. (C) Transmittance on resonance as
a function of incident power. (Inset) A weak, resonant coherent state is reflected by the atom.}
\end{figure}

The electromagnetic field in the transmission line is described by
an incoming voltage wave $V_{in}$, a reflected wave $V_{R}$ and
a transmitted wave $V_{T}$. In Fig.1A, the transmittance is
defined as $T=|V_{T}/V_{in}|^2$. For a weak coherent drive on resonance
with the atom, we expect to see full reflection of the incident
signal \cite{Shen2,Chang}. This can be understood in terms of interference between
the incident wave and wave scattered from the atom, which
destructively interferes in transmission and constructively interferes
in reflection \cite{Shen2,Chang}. In the sample measured here, we achieved extinction of more than 99\% in transmittance, as shown in Fig.\,1C. By measuring the transmission
coefficient as a function of probe frequency and probe power $P$, we extract
$\omega_{01}/2\pi=5.12$ GHz, $\Gamma_{10}/2\pi=41$ MHz,
$\Gamma_{\phi}/2\pi=1$ MHz. The relaxation rate
$\Gamma_{10}$ is dominated by the coupling to the line and is much greater than the pure dephasing $\Gamma_{\phi}$ in
our system. We define an average number of
photons per interaction time $2\pi/\Gamma_{10}$ as $N\equiv 2\pi
P/(\hbar\omega_p\Gamma_{10})$. $N = 1$ for a power of $-128$ dBm.

The resonant electromagnetic field reflected from or
transmitted through the atom (depending on the choice of input port,
see Fig 1B) is fed through two circulators to a commercial $90^o$
hybrid coupler, with its other input terminated with 50
$\Omega$. The hybrid coupler effectively acts as a microwave beam
splitter. Ideally, the signal coming into the other input of the hybrid
coupler should be vacuum. The two outputs of the beam
splitter are sent to two nominally identical microwave HEMT amplifiers at 4.2 K that have system noise temperatures of 7 K. We make the essential
assumption in our analysis that the noise added by the two
amplifiers is uncorrelated. After further amplification, the two outputs are fed into a pair of vector digitizers, which capture the voltage amplitude. We then can choose to digitally filter the voltage data to a desired bandwidth, BW. This setup enables us to perform Hanbury Brown-Twiss (HBT) \cite{Brown,Gabelli,Silva,Grosse,Menzel}
measurements using linear quadrature detectors, \textit{i.e.}, amplifiers.

 \begin{figure*}
 \includegraphics[width=2\columnwidth]{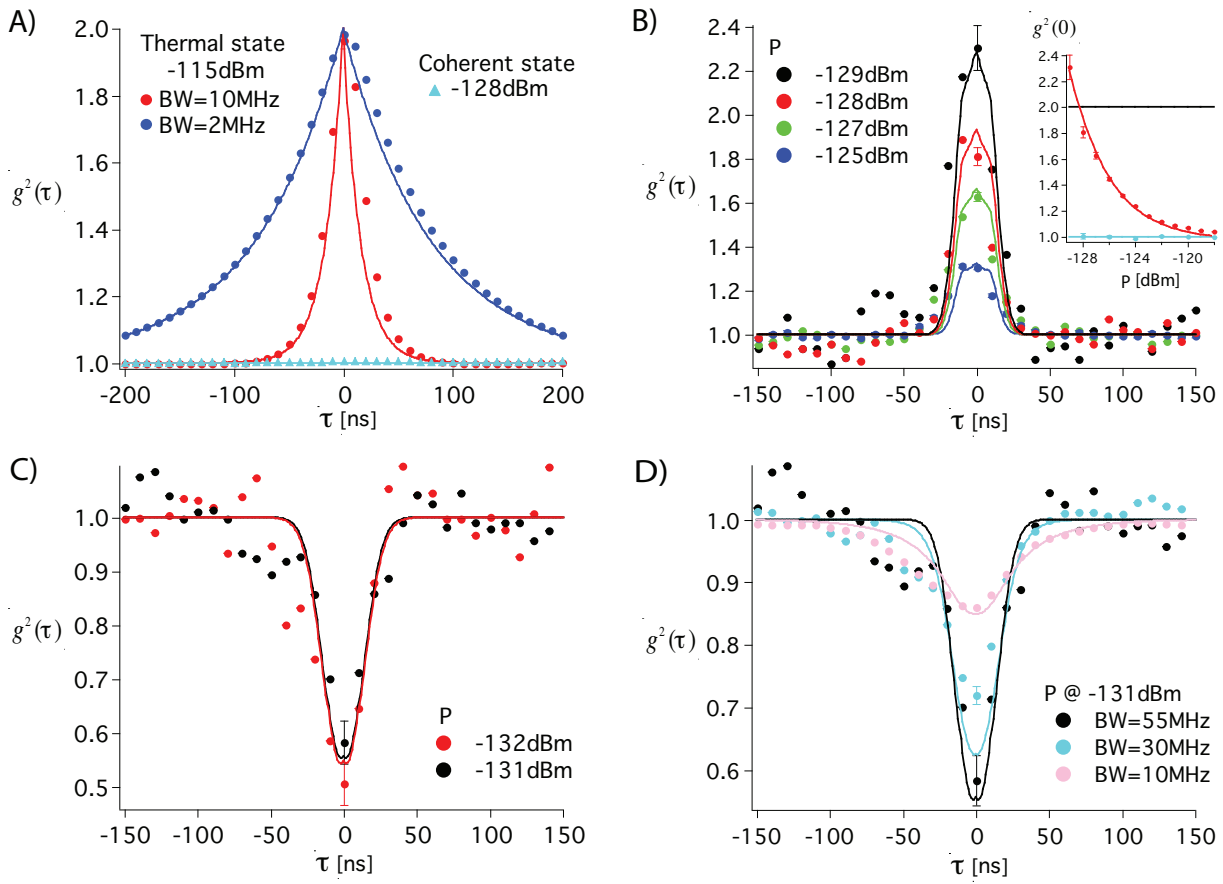}
\caption{ Second-order correlation function of a thermal state, a coherent state and the states generated by the artificial atom. A) $g^{(2)}$ of a thermal state and a coherent state
as a function of delay time $\tau$. B) $g^{(2)}$ of the resonant transmitted microwaves as a function of delay time for four different
incident powers. Inset: $g^{(2)}(0)$ as a
function of incident power. For comparison, the result for a thermal state and
a coherent state are also plotted. We see that the transmitted field
statistics(red curve) approaches that of a coherent field at high incident
power, as expected. For a coherent
state $g^{(2)}(0)=1$ is independent of incident power (blue). C) $g^{(2)}$ of a resonant reflected field as a
function of delay time for two different incident powers. The
antibunching behavior reveals the quantum nature of the field. The
curves shown here had a digital filter with a 55 MHz bandwidth
applied to each detector. D) $g^{(2)}$ of a resonant reflected field as a function of delay time at -131dBm incident power for different filter bandwidths. As the bandwidth decreases, the antibunching dip vanishes. The solid curves in B), C) and D) are the theory curves, including trigger jitter model (see text). We extract a temperature of 50 mK from these fits, with no additional free fitting parameters. The peculiar feature of $g^{(2)}$ around zero in the theory curves in B) is due to the trigger jitter model. The error bar indicated for each data set is the same for all the points.}
\end{figure*}

Second-order correlation measurements provide a method to 
characterize microwave states generated in our system. In
particular, they provide a statistical tool to show that the scattered light is nonclassical. For reference, we first considered a
thermal state and a coherent state. The thermal state \footnote{Technically, it is a chaotic state.} was
generated by simply amplifying the noise of a 50 $\Omega$ resistor through room
temperature amplifiers before sending the signal down through the
transmission line in our set-up, with the qubit off-resonance. The two digitizers were set to a sampling
frequency of $10^8$ samples/sec and then the voltages were digitally
filtered separately. Finally, we determine the power-power correlations as a function of delay time between the two outputs. 

The second order correlation function can be expressed as
\begin{eqnarray*}
g^{(2)}(\tau)=1+\frac{\left \langle \Delta P_1(t)\Delta P_2(t+\tau)\right \rangle}{[\left \langle P_1(t) \right \rangle - \left \langle P_{1,N}(t) \right \rangle][\left \langle P_2(t)\right \rangle-\left \langle P_{2,N}(t) \right \rangle]},
\end{eqnarray*}
where $\tau$ is the delay time between the two digitizers, $P_1,P_2$
are the output powers in ports 1 and 2, respectively.  $P_{1,N},P_{2,N}$
are the amplifier noise in ports 1 and 2, respectively, when the incident source is off. Therefore, $[\left \langle P_i(t)\right \rangle-\left \langle P_{i,N}(t) \right \rangle]$ represents the net power of the field from output port i, where $i=1,2$. 
$\left<\Delta P_1\Delta P_2\right>$ is the covariance of the output powers in port 1 and 2, defined as
$\left<(P_1- \left< P_1\right>)(P_2- \left< P_2\right>)\right>$. The following assumptions were made: (1) the amplifier noise originating from the two independent detection chains is
uncorrelated, and (2) the 50 $\Omega$ terminator is in its ground state (5 GHz photons at $\sim$50 mK). In Fig. 2A, we
show $g^{(2)}$ as a function of delay time $\tau$, for a thermal state
with two different filter bandwidths, and also for a coherent state.
For thermal states, $g^{(2)}(0)=2$ regardless of the filter bandwidth BW. The width of $g^{(2)}(\tau)$ for the thermal state is
determined by the filter function. For the filter used
$g^{(2)}(\tau)=1+e^{-2\pi BW|\tau|}$, the solid curves of the thermal
state in Fig. 2A show this equation with no free fitting parameters.  
We had a trigger jitter of $\pm1$ sample between the two
digitizers. To minimize the effect of this trigger jitter, we oversample and then digitally filter (average) the data in all the  $g^{(2)}$ measurements. For a coherent
state, we expect $g^{(2)}(\tau)=1$. This is indeed what we find if
our atom is off-resonance from our applied coherent source.

After making these initial measurements, we made second order
correlation measurements to study the field transmitted through our
atom. We applied an on-resonance microwave drive and
measured $g^{(2)}(\tau)$  in transmission for different incident powers as shown in
Fig. 2B. Here the sampling frequency was again set to $10^8$ samples/s
with BW=55 MHz. At the lowest power ($P=-129$ dBm, $N=0.8$) that
we can readily measure, we see superbunching of the photons as expected \cite{Chang}, with $g^{(2)}(\tau=0)=2.31\pm 0.09>2$. Superbunching occurs because the one-photon state of the incident field has been selectively filtered and reflected while the two-photon state is more likely transmitted (the three-photon components and higher are negligible). This transmitted state generated from our qubit is thus bunched even more than a thermal state. For high powers, where $N\gg1$, we find
$g^{(2)}(\tau)=1$ as expected, because most of the coherent signal then passes
through the transmission line without interacting with the atom owing to saturation of the atomic response. The correlation once again looks like a coherent state. For
all measurements shown here we find, $g^{(2)}(\infty)=1$, as
expected. In the right inset of Fig. 2B, we plot $g^{(2)}(0)$ as a function of incident
power, and clearly see the bunching behavior decrease as the
incident power increases. For comparison, we also plot the coherent state and thermal state. 

In Fig. 2C, we show the measured $g^{(2)}(\tau)$ of the reflected
field from our atom. At low powers, where $N\ll1$, we clearly observe antibunching of the field \cite{Chang}. Each trace here was
collected, computed and averaged over 17 hours, corresponding to
$2.4\times10^{11}$ measured quadrature field samples (2 Tbyte
of data). The antibunching behavior at the lowest power($P=-132$ dBm, $N=0.4$), $g^{(2)}(0)=0.51\pm 0.05$, reveals the quantum nature of the
field. Ideally, we would find $g^{(2)}(0)=0$ as the atom can only absorb
and emit one photon at a time. The non-zero $g^{(2)}(0)$ we measured is most likely from the combination of trigger jitter
between the two digitizers and the filter bandwidth (here 55 MHz).
In Fig. 2D, we clearly see how the antibunching dip depends on bandwidth. For small BW, where BW $\ll\Gamma_{01}$, the
antibunching dip vanishes entirely. With a small BW or long
sampling time, the time dynamics of antibunching cannot be resolved.
In other words, within that time interval, the atom has
had time to absorb and emit many photons. 

To obtain the theoretical results in Fig. 2 B, C, and D, we use a master equation formalism. The digital filter is modeled by a single-mode resonator. A master equation describing both the transmon and the resonator is derived using the formalism of cascaded quantum systems. To model the effect of the trigger jitter, the value of $g^{(2)}(\tau)$ at each point is replaced by the average value of $g^{(2)}(\tau)$, $g^{(2)}(\tau$-10 ns) and $g^{(2)}(\tau$+10 ns).

Our artifical atom selectively filters out the Fock state $n=1$ from the
input coherent state.  As a result, the reflected and transmitted
field display antibunched and superbunched statistics, respectively.  Thus, the
qubit acts as a passive photon-number filter, converting a coherent
microwave state to a nonclassical one, with high production rate.

\begin{figure}
 \includegraphics[width = 3in]{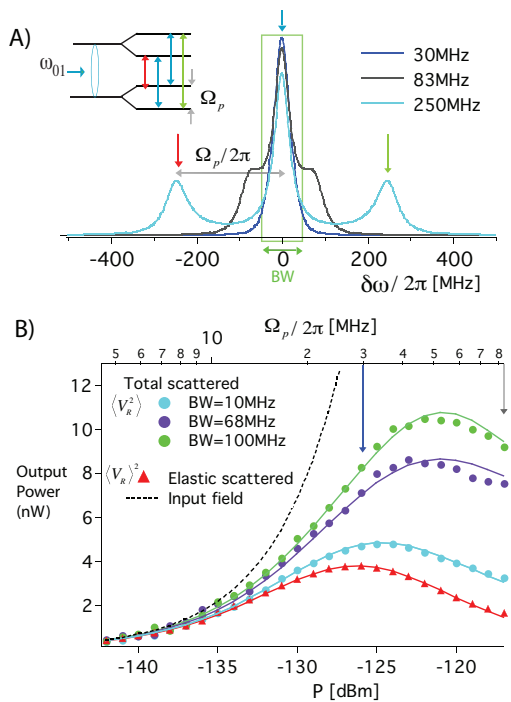}
\caption{Elastic vs. inelastic scattering from the artificial atom. A) A microwave pump is
applied at $\omega_{01}$.  As the power of the $\omega_{01}$ pump
increases, the Mollow triplet appears in the spectrum with peak
separation equal to the Rabi frequency $\Omega_{p}$. (inset) Dressed
state picture of the energy levels. B) The coherently/elastically
reflected power (phase-sensitive average, red curve) or total reflected
power (phase-insensitive average, green, grey and blue curves) as a function of resonant
incident powers for different bandwidths (BW). The total power
reflected is the sum of both the elastic and inelastic fields,
after we subtract off the amplifier noise power.  Solid curves are the theory fits to experimental data. The black dash curve shows the input power for comparison. At low powers, $N<<1$, we observed that
the reflected field is mostly first-order coherent: the elastically
reflected power is the same as the total reflected power. At high
powers, $N>1$, more and more photons are inelastically scattered as
the Mollow triplet begins to emerge. The wider the measurement
bandwidth, the more of the Mollow triplet we capture. Note that the output power includes the 79 dB gain of the amplifiers.}
\end{figure}

While the scattered field requires a purely quantum description,
it can still maintain first-order coherence similar to a classical field, as shown below. We can define the first-order correlation function in steady state as $g^{(1)}=\left \langle V\right\rangle^2/\left \langle V^2\right \rangle$. First-order coherence then refers to $g^{(1)}=1$. As expected, for a
thermal source, this function is 0 and for a coherent
state it is 1.

The first-order coherence properties of the scattered resonance
field strongly depend on the Rabi frequency $\Omega_{p}$ and the
relaxation rate $\Gamma_{10}$ of the atom. The Rabi frequency
$\Omega_{p}$ is linearly proportional to the amplitude of the drive,
$\Omega_{p}\propto\sqrt{P}$ \cite{Hoi}.  For $\Omega_{p}\ll\Gamma_{10}$, we expect the scattered fields to be
mostly coherent to first order, with a spectrum well described by a
delta function \cite{Loudon} at $\omega_{01}$. For $\Omega_{p}>\Gamma_{10}$, the scattered fields contain
three additional inelastic Lorentzians (known as the Mollow Triplet
\cite{Mollow}) centered on the frequencies
$\omega=\omega_{01}$, $\omega=\omega_{01}+\Omega_p$ and
$\omega=\omega_{01}-\Omega_p$, indicated in the theory plot Fig.\,3A. 

We send a single tone at $\omega_{01}$ and measure these scattered (reflected) fields from only one of the output ports, as shown in Fig.\,3B. Note that we see the same behavior from both outputs. We use a phase-sensitive average $\left \langle V\right \rangle^2$ to capture
the elastic (coherent) component of the scattered field. For the total
scattered field, the sum of the elastically and inelastically scattered fields, we use a phase-insensitive average $\left \langle
V^2\right \rangle$. The amount of the inelastic field that we
capture depends on the bandwidth (BW) of our measurement, as indicated in
Fig.\,3A. The solid curves are the theory fits, using the model in Fig.\,3A (integrating the Mollow triplet), with no free fitting parameters (using the same parameters
extracted before). As expected, at low incident power, the total
scattered field is roughly equal to the elastically scattered (coherent)
field. This indicates that the field is first-order coherent with $g^{(1)}\simeq1$. We note that this is also the regime where antibunching is observed. At high incident fields, where $\Omega_p> \Gamma_{10}$, the main contribution
to the total field is from inelastic scattering.  

In conclusion, we investigated the scattering properties of a single
artificial atom in a 1D open space by measuring the first, second and fourth
moments of the voltage field. We verified the quantum nature of the
scattered field using the second-order correlation function, while also
showing that the field maintained first-order coherence. In fact, the
whole process leads to a redistribution of the photon number state \cite{Zheng}.

We plan to investigate applications of this phenomenon, such as generating single-photon
states on demand.  This system may offer advantages over placing an
artificial atom in a cavity \cite{Bozyigit,Lang,Houck1}, for instance. In our system, regardless of the
strength of the input pulse only a single photon state is reflected. This
eliminates the need for precise Rabi pulses to excite the atom as in a
cavity-based photon source. Furthermore, our generated single photons can
maintain the same envelope as the incident coherent state with a wide
bandwidth limited by the atom relaxation rate. For a cavity-based photon
source, the bandwidth is limited by the cavity width and subject to the
problem of stochastic release by the cavity.

We acknowledge financial support from the Swedish Research Council, the Wallenberg foundation, SOLID and from the European Research Council. We would also like to acknowledge G. Milburn and T. Stace for fruitful discussions.


\begin{thebibliography}{10}%
\makeatletter
\providecommand \@ifxundefined [1]{%
 \ifx #1\undefined \expandafter \@firstoftwo
 \else \expandafter \@secondoftwo
\fi
}%
\providecommand \@ifnum [1]{%
 \ifnum #1\expandafter \@firstoftwo
 \else \expandafter \@secondoftwo
\fi
}%
\providecommand \enquote [1]{``#1''}%
\providecommand \bibnamefont  [1]{#1}%
\providecommand \bibfnamefont [1]{#1}%
\providecommand \citenamefont [1]{#1}%
\providecommand\href[0]{\@sanitize\@href}%
\providecommand\@href[1]{\endgroup\@@startlink{#1}\endgroup\@@href}%
\providecommand\@@href[1]{#1\@@endlink}%
\providecommand \@sanitize [0]{\begingroup\catcode`\&12\catcode`\#12\relax}%
\@ifxundefined \pdfoutput {\@firstoftwo}{%
 \@ifnum{\z@=\pdfoutput}{\@firstoftwo}{\@secondoftwo}%
}{%
 \providecommand\@@startlink[1]{\leavevmode}%
 \providecommand\@@endlink[0]{}%
}{%
 \providecommand\@@startlink[1]{%
  \leavevmode
  \pdfstartlink
   attr{/Border[0 0 1 ]/H/I/C[0 1 1]}%
   user{/Subtype/Link/A<</Type/Action/S/URI/URI(#1)>>}%
  \relax
 }%
 \providecommand\@@endlink[0]{\pdfendlink}%
}%
\providecommand \url  [0]{\begingroup\@sanitize \@url }%
\providecommand \@url [1]{\endgroup\@href {#1}{\urlprefix}}%
\providecommand \urlprefix [0]{URL }%
\providecommand \Eprint[0]{\href }%
\@ifxundefined \urlstyle {%
  \providecommand \doi [1]{doi:\discretionary{}{}{}#1}%
}{%
  \providecommand \doi [0]{doi:\discretionary{}{}{}\begingroup
  \urlstyle{rm}\Url }%
}%
\providecommand \doibase [0]{http://dx.doi.org/}%
\providecommand \Doi[1]{\href{\doibase#1}}%
\providecommand \bibAnnote [3]{%
  \BibitemShut{#1}%
  \begin{quotation}\noindent
    \textsc{Key:}\ #2\\\textsc{Annotation:}\ #3%
  \end{quotation}%
}%
\providecommand \bibAnnoteFile [2]{%
  \IfFileExists{#2}{\bibAnnote {#1} {#2} {\input{#2}}}{}%
}%
\providecommand \typeout [0]{\immediate \write \m@ne }%
\providecommand \selectlanguage [0]{\@gobble}%
\providecommand \bibinfo [0]{\@secondoftwo}%
\providecommand \bibfield [0]{\@secondoftwo}%
\providecommand \translation [1]{[#1]}%
\providecommand \BibitemOpen[0]{}%
\providecommand \bibitemStop [0]{}%
\providecommand \bibitemNoStop [0]{.\EOS\space}%
\providecommand \EOS [0]{\spacefactor3000\relax}%
\providecommand \BibitemShut [1]{\csname bibitem#1\endcsname}%
\bibitem{Astafiev1}%
  \BibitemOpen
  \bibfield{author}{%
  \bibinfo {author} {\bibfnamefont{O.}~\bibnamefont{Astafiev}}, \bibinfo
  {author} {\bibfnamefont{A.~M.}\ \bibnamefont{Zagoskin}}, \bibinfo {author}
  {\bibfnamefont{Jr.}~\bibnamefont{A.~A.~Abdumalikov}}, \bibinfo {author}
  {\bibfnamefont{Y.~A.}\ \bibnamefont{Pashkin}}, \bibinfo {author}
  {\bibfnamefont{T.}~\bibnamefont{Yamamoto}}, \bibinfo {author}
  {\bibfnamefont{K.}~\bibnamefont{Inomata}}, \bibinfo {author}
  {\bibfnamefont{Y.}~\bibnamefont{Nakamura}},\ and\ \bibinfo {author}
  {\bibfnamefont{J.~S.}\ \bibnamefont{Tsai}},\ }%
  \bibfield{journal}{%
  \bibinfo {journal} {Science}\ }%
  \textbf{\bibinfo {volume} {327}},\ \bibinfo {pages} {840} (\bibinfo {year}
  {2010})%
  \bibAnnoteFile{NoStop}{Astafiev1}%
\bibitem{Hoi}%
  \BibitemOpen
  \bibfield{author}{%
  \bibinfo {author} {\bibfnamefont{I.~C.}\ \bibnamefont{Hoi}}, \bibinfo
  {author} {\bibfnamefont{C.~M.}\ \bibnamefont{Wilson}}, \bibinfo {author}
  {\bibfnamefont{G.}~\bibnamefont{Johansson}}, \bibinfo {author}
  {\bibfnamefont{T.}~\bibnamefont{Palomaki}}, \bibinfo {author}
  {\bibfnamefont{B.}~\bibnamefont{Peropadre}},\ and\ \bibinfo {author}
  {\bibfnamefont{P.}~\bibnamefont{Delsing}},\ }%
  \bibfield{journal}{%
  \bibinfo {journal} {Phys. Rev. Lett.}\ }%
  \textbf{\bibinfo {volume} {107}},\ \bibinfo {pages} {073601} (\bibinfo {year}
  {2011})%
  \bibAnnoteFile{NoStop}{Hoi}%
\bibitem{Zumofen}%
  \BibitemOpen
  \bibfield{author}{%
  \bibinfo {author} {\bibfnamefont{G.}~\bibnamefont{Zumofen}}, \bibinfo
  {author} {\bibfnamefont{N.~M.}\ \bibnamefont{Mojarad}}, \bibinfo {author}
  {\bibfnamefont{V.}~\bibnamefont{Sandoghdar}},\ and\ \bibinfo {author}
  {\bibfnamefont{M.}~\bibnamefont{Agio}},\ }%
  \bibfield{journal}{%
  \bibinfo {journal} {Physical Review Letters}\ }%
  \textbf{\bibinfo {volume} {101}},\ \bibinfo {pages} {180404} (\bibinfo {year}
  {2008})%
  \bibAnnoteFile{NoStop}{Zumofen}%
\bibitem{Chang}%
  \BibitemOpen
  \bibfield{author}{%
  \bibinfo {author} {\bibfnamefont{D.~E.}\ \bibnamefont{Chang}}, \bibinfo
  {author} {\bibfnamefont{A.~S.}\ \bibnamefont{Sorensen}}, \bibinfo {author}
  {\bibfnamefont{E.~A.}\ \bibnamefont{Demler}},\ and\ \bibinfo {author}
  {\bibfnamefont{M.~D.}\ \bibnamefont{Lukin}},\ }%
  \bibfield{journal}{%
  \bibinfo {journal} {Nature Physics}\ }%
  \textbf{\bibinfo {volume} {3}},\ \bibinfo {pages} {807} (\bibinfo {year}
  {2007})%
  \bibAnnoteFile{NoStop}{Chang}%
\bibitem{Loudon}%
  \BibitemOpen
  \bibfield{author}{%
  \bibinfo {author} {\bibfnamefont{R.}~\bibnamefont{Loudon}},\ }%
  \emph{\bibinfo {title} {The Quantum Theory of Light}}\ (\bibinfo {publisher}
  {Oxford science publications},\ \bibinfo {year} {2010})%
  \bibAnnoteFile{NoStop}{Loudon}%
\bibitem{Zheng}%
  \BibitemOpen
  \bibfield{author}{%
  \bibinfo {author} {\bibfnamefont{H.}~\bibnamefont{Zheng}}, \bibinfo {author}
  {\bibfnamefont{D.~J.}\ \bibnamefont{Gauthier}},\ and\ \bibinfo {author}
  {\bibfnamefont{H.~U.}\ \bibnamefont{Baranger}},\ }%
  \bibfield{journal}{%
  \bibinfo {journal} {Physical Review A}\ }%
  \textbf{\bibinfo {volume} {82}},\ \bibinfo {pages} {063816} (\bibinfo {year}
  {2010})%
  \bibAnnoteFile{NoStop}{Zheng}%
\bibitem{chris}%
  \BibitemOpen
  \bibfield{author}{%
  \bibinfo {author} {\bibfnamefont{C.~M.}\ \bibnamefont{Wilson}}, \bibinfo
  {author} {\bibfnamefont{G.}~\bibnamefont{Johansson}}, \bibinfo {author}
  {\bibfnamefont{A.}~\bibnamefont{Pourkabirian}}, \bibinfo {author}
  {\bibfnamefont{M.}~\bibnamefont{Simoen}}, \bibinfo {author}
  {\bibfnamefont{J.~R.}\ \bibnamefont{Johansson}}, \bibinfo {author}
  {\bibfnamefont{T.}~\bibnamefont{Duty}}, \bibinfo {author}
  {\bibfnamefont{F.}~\bibnamefont{Nori}},\ and\ \bibinfo {author}
  {\bibfnamefont{P.}~\bibnamefont{Delsing}},\ }%
  \bibfield{journal}{%
  \bibinfo {journal} {Nature}\ }%
  \textbf{\bibinfo {volume} {479}},\ \bibinfo {pages} {376} (\bibinfo {year}
  {2011})%
  \bibAnnoteFile{NoStop}{chris}%
\bibitem{Eichler}%
  \BibitemOpen
  \bibfield{author}{%
  \bibinfo {author} {\bibfnamefont{C.}~\bibnamefont{Eichler}}, \bibinfo
  {author} {\bibfnamefont{D.}~\bibnamefont{Bozyigit}}, \bibinfo {author}
  {\bibfnamefont{C.}~\bibnamefont{Lang}}, \bibinfo {author}
  {\bibfnamefont{M.}~\bibnamefont{Baur}}, \bibinfo {author}
  {\bibfnamefont{L.}~\bibnamefont{Steffen}}, \bibinfo {author}
  {\bibfnamefont{J.~M.}\ \bibnamefont{Fink}}, \bibinfo {author}
  {\bibfnamefont{S.}~\bibnamefont{Filipp}},\ and\ \bibinfo {author}
  {\bibfnamefont{A.}~\bibnamefont{Wallraff}},\ }%
  \bibfield{journal}{%
  \Doi{10.1103/PhysRevLett.107.113601}{\bibinfo {journal} {Physical Review
  Letters}}\ }%
  \textbf{\bibinfo {volume} {107}},\ \bibinfo {pages} {113601} (\bibinfo {year}
  {2011})%
  \bibAnnoteFile{NoStop}{Eichler}%
\bibitem{Eichler2}%
  \BibitemOpen
  \bibfield{author}{%
  \bibinfo {author} {\bibfnamefont{C.}~\bibnamefont{Eichler}}, \bibinfo
  {author} {\bibfnamefont{D.}~\bibnamefont{Bozyigit}}, \bibinfo {author}
  {\bibfnamefont{C.}~\bibnamefont{Lang}}, \bibinfo {author}
  {\bibfnamefont{L.}~\bibnamefont{Steffen}}, \bibinfo {author}
  {\bibfnamefont{J.}~\bibnamefont{Fink}},\ and\ \bibinfo {author}
  {\bibfnamefont{A.}~\bibnamefont{Wallraff}},\ }%
  \bibfield{journal}{%
  \Doi{10.1103/PhysRevLett.106.220503}{\bibinfo {journal} {Physical Review
  Letters}}\ }%
  \textbf{\bibinfo {volume} {106}},\ \bibinfo {pages} {220503} (\bibinfo {year}
  {2011})%
  \bibAnnoteFile{NoStop}{Eichler2}%
\bibitem{Reulet}%
  \BibitemOpen
  \bibfield{author}{%
  \bibinfo {author} {\bibfnamefont{B.}~\bibnamefont{Reulet}}, \bibinfo {author}
  {\bibfnamefont{J.}~\bibnamefont{Senzier}},\ and\ \bibinfo {author}
  {\bibfnamefont{D.~E.}\ \bibnamefont{Prober}},\ }%
  \bibfield{journal}{%
  \Doi{10.1103/PhysRevLett.91.196601}{\bibinfo {journal} {Physical Review
  Letters}}\ }%
  \textbf{\bibinfo {volume} {91}},\ \bibinfo {pages} {196601} (\bibinfo {year}
  {2003})%
  \bibAnnoteFile{NoStop}{Reulet}%
\bibitem{Mallet}%
  \BibitemOpen
  \bibfield{author}{%
  \bibinfo {author} {\bibfnamefont{F.}~\bibnamefont{Mallet}}, \bibinfo {author}
  {\bibfnamefont{M.~A.}\ \bibnamefont{Castellanos-Beltran}}, \bibinfo {author}
  {\bibfnamefont{H.~S.}\ \bibnamefont{Ku}}, \bibinfo {author}
  {\bibfnamefont{S.}~\bibnamefont{Glancy}}, \bibinfo {author}
  {\bibfnamefont{E.}~\bibnamefont{Knill}}, \bibinfo {author}
  {\bibfnamefont{K.~D.}\ \bibnamefont{Irwin}}, \bibinfo {author}
  {\bibfnamefont{G.~C.}\ \bibnamefont{Hilton}}, \bibinfo {author}
  {\bibfnamefont{L.~R.}\ \bibnamefont{Vale}},\ and\ \bibinfo {author}
  {\bibfnamefont{K.~W.}\ \bibnamefont{Lehnert}},\ }%
  \bibfield{journal}{%
  \Doi{10.1103/PhysRevLett.106.220502}{\bibinfo {journal} {Physical Review
  Letters}}\ }%
  \textbf{\bibinfo {volume} {106}},\ \bibinfo {pages} {220502} (\bibinfo {year}
  {2011})%
  \bibAnnoteFile{NoStop}{Mallet}%
\bibitem{Koch}%
  \BibitemOpen
  \bibfield{author}{%
  \bibinfo {author} {\bibfnamefont{J.}~\bibnamefont{Koch}}, \bibinfo {author}
  {\bibfnamefont{T.~M.}\ \bibnamefont{Yu}}, \bibinfo {author}
  {\bibfnamefont{J.}~\bibnamefont{Gambetta}}, \bibinfo {author}
  {\bibfnamefont{A.~A.}\ \bibnamefont{Houck}}, \bibinfo {author}
  {\bibfnamefont{D.~I.}\ \bibnamefont{Schuster}}, \bibinfo {author}
  {\bibfnamefont{J.}~\bibnamefont{Majer}}, \bibinfo {author}
  {\bibfnamefont{A.}~\bibnamefont{Blais}}, \bibinfo {author}
  {\bibfnamefont{M.~H.}\ \bibnamefont{Devoret}}, \bibinfo {author}
  {\bibfnamefont{S.~M.}\ \bibnamefont{Girvin}},\ and\ \bibinfo {author}
  {\bibfnamefont{R.~J.}\ \bibnamefont{Schoelkopf}},\ }%
  \bibfield{journal}{%
  \bibinfo {journal} {Physical Review A}\ }%
  \textbf{\bibinfo {volume} {76}},\ \bibinfo {pages} {042319} (\bibinfo {year}
  {2007})%
  \bibAnnoteFile{NoStop}{Koch}%
\bibitem{Shen2}%
  \BibitemOpen
  \bibfield{author}{%
  \bibinfo {author} {\bibfnamefont{J.~T.}\ \bibnamefont{Shen}}\ and\ \bibinfo
  {author} {\bibfnamefont{S.~H.}\ \bibnamefont{Fan}},\ }%
  \bibfield{journal}{%
  \bibinfo {journal} {Physical Review Letters}\ }%
  \textbf{\bibinfo {volume} {95}},\ \bibinfo {pages} {213001} (\bibinfo {year}
  {2005})%
  \bibAnnoteFile{NoStop}{Shen2}%
\bibitem{Brown}%
  \BibitemOpen
  \bibfield{author}{%
  \bibinfo {author} {\bibfnamefont{R.~H.}\ \bibnamefont{Brown}}\ and\ \bibinfo
  {author} {\bibfnamefont{R.~Q.}\ \bibnamefont{Twiss}},\ }%
  \bibfield{journal}{%
  \bibinfo {journal} {Nature}\ }%
  \textbf{\bibinfo {volume} {177}},\ \bibinfo {pages} {27} (\bibinfo {year}
  {1956})%
  \bibAnnoteFile{NoStop}{Brown}%
\bibitem{Gabelli}%
  \BibitemOpen
  \bibfield{author}{%
  \bibinfo {author} {\bibfnamefont{J.}~\bibnamefont{Gabelli}}, \bibinfo
  {author} {\bibfnamefont{L.-H.}\ \bibnamefont{Reydellet}}, \bibinfo {author}
  {\bibfnamefont{G.}~\bibnamefont{F{\`e}ve}}, \bibinfo {author}
  {\bibfnamefont{J.-M.}\ \bibnamefont{Berroir}}, \bibinfo {author}
  {\bibfnamefont{B.}~\bibnamefont{Pla{\c c}ais}}, \bibinfo {author}
  {\bibfnamefont{P.}~\bibnamefont{Roche}},\ and\ \bibinfo {author}
  {\bibfnamefont{D.~C.}\ \bibnamefont{Glattli}},\ }%
  \bibfield{journal}{%
  \bibinfo {journal} {Phys. Rev. Lett.}\ }%
  \textbf{\bibinfo {volume} {93}},\ \bibinfo {pages} {056801} (\bibinfo {year}
  {2004})%
  \bibAnnoteFile{NoStop}{Gabelli}%
\bibitem{Silva}%
  \BibitemOpen
  \bibfield{author}{%
  \bibinfo {author} {\bibfnamefont{M.~P.}\ \bibnamefont{da~Silva}}, \bibinfo
  {author} {\bibfnamefont{D.}~\bibnamefont{Bozyigit}}, \bibinfo {author}
  {\bibfnamefont{A.}~\bibnamefont{Wallraff}},\ and\ \bibinfo {author}
  {\bibfnamefont{A.}~\bibnamefont{Blais}},\ }%
  \bibfield{journal}{%
  \bibinfo {journal} {Phys. Rev. A}\ }%
  \textbf{\bibinfo {volume} {82}},\ \bibinfo {pages} {043804} (\bibinfo {year}
  {2010})%
  \bibAnnoteFile{NoStop}{Silva}%
\bibitem{Grosse}%
  \BibitemOpen
  \bibfield{author}{%
  \bibinfo {author} {\bibfnamefont{N.~B.}\ \bibnamefont{Grosse}}, \bibinfo
  {author} {\bibfnamefont{T.}~\bibnamefont{Symul}}, \bibinfo {author}
  {\bibfnamefont{M.}~\bibnamefont{Stobi{\'n}ska}}, \bibinfo {author}
  {\bibfnamefont{T.~C.}\ \bibnamefont{Ralph}},\ and\ \bibinfo {author}
  {\bibfnamefont{P.~K.}\ \bibnamefont{Lam}},\ }%
  \bibfield{journal}{%
  \bibinfo {journal} {Phys. Rev. Lett.}\ }%
  \textbf{\bibinfo {volume} {98}},\ \bibinfo {pages} {153603} (\bibinfo {year}
  {2007})%
  \bibAnnoteFile{NoStop}{Grosse}%
\bibitem{Menzel}%
  \BibitemOpen
  \bibfield{author}{%
  \bibinfo {author} {\bibfnamefont{E.~P.}\ \bibnamefont{Menzel}}, \bibinfo
  {author} {\bibfnamefont{F.}~\bibnamefont{Deppe}}, \bibinfo {author}
  {\bibfnamefont{M.}~\bibnamefont{Mariantoni}}, \bibinfo {author}
  {\bibfnamefont{M.~{\'A}.}\ \bibnamefont{Araque Caballero}}, \bibinfo {author}
  {\bibfnamefont{A.}~\bibnamefont{Baust}}, \bibinfo {author}
  {\bibfnamefont{T.}~\bibnamefont{Niemczyk}}, \bibinfo {author}
  {\bibfnamefont{E.}~\bibnamefont{Hoffmann}}, \bibinfo {author}
  {\bibfnamefont{A.}~\bibnamefont{Marx}}, \bibinfo {author}
  {\bibfnamefont{E.}~\bibnamefont{Solano}},\ and\ \bibinfo {author}
  {\bibfnamefont{R.}~\bibnamefont{Gross}},\ }%
  \bibfield{journal}{%
  \bibinfo {journal} {Phys. Rev. Lett.}\ }%
  \textbf{\bibinfo {volume} {105}},\ \bibinfo {pages} {100401} (\bibinfo {year}
  {2010})%
  \bibAnnoteFile{NoStop}{Menzel}%
\bibitem{Note1}%
  \BibitemOpen
  \bibinfo {note} {Technically, it is a chaotic state.}%
  \bibAnnoteFile{Stop}{Note1}%
\bibitem{Mollow}%
  \BibitemOpen
  \bibfield{author}{%
  \bibinfo {author} {\bibfnamefont{B.~R.}\ \bibnamefont{Mollow}},\ }%
  \bibfield{journal}{%
  \bibinfo {journal} {Physical Review}\ }%
  \textbf{\bibinfo {volume} {188}},\ \bibinfo {pages} {1969} (\bibinfo {year}
  {1969})%
  \bibAnnoteFile{NoStop}{Mollow}%
\bibitem{Bozyigit}%
  \BibitemOpen
  \bibfield{author}{%
  \bibinfo {author} {\bibfnamefont{D.}~\bibnamefont{Bozyigit}}, \bibinfo
  {author} {\bibfnamefont{C.}~\bibnamefont{Lang}}, \bibinfo {author}
  {\bibfnamefont{L.}~\bibnamefont{Steffen}}, \bibinfo {author}
  {\bibfnamefont{J.~M.}\ \bibnamefont{Fink}}, \bibinfo {author}
  {\bibfnamefont{C.}~\bibnamefont{Eichler}}, \bibinfo {author}
  {\bibfnamefont{M.}~\bibnamefont{Baur}}, \bibinfo {author}
  {\bibfnamefont{R.}~\bibnamefont{Bianchetti}}, \bibinfo {author}
  {\bibfnamefont{P.~J.}\ \bibnamefont{Leek}}, \bibinfo {author}
  {\bibfnamefont{S.}~\bibnamefont{Filipp}}, \bibinfo {author}
  {\bibfnamefont{M.~P.}\ \bibnamefont{da~Silva}}, \bibinfo {author}
  {\bibfnamefont{A.}~\bibnamefont{Blais}},\ and\ \bibinfo {author}
  {\bibfnamefont{A.}~\bibnamefont{Wallraff}},\ }%
  \bibfield{journal}{%
  \bibinfo {journal} {Nature Physics}\ }%
  \textbf{\bibinfo {volume} {7}},\ \bibinfo {pages} {154} (\bibinfo {year}
  {2011})%
  \bibAnnoteFile{NoStop}{Bozyigit}%
\bibitem{Lang}%
  \BibitemOpen
  \bibfield{author}{%
  \bibinfo {author} {\bibfnamefont{C.}~\bibnamefont{Lang}}, \bibinfo {author}
  {\bibfnamefont{D.}~\bibnamefont{Bozyigit}}, \bibinfo {author}
  {\bibfnamefont{C.}~\bibnamefont{Eichler}}, \bibinfo {author}
  {\bibfnamefont{L.}~\bibnamefont{Steffen}}, \bibinfo {author}
  {\bibfnamefont{J.~M.}\ \bibnamefont{Fink}}, \bibinfo {author}
  {\bibfnamefont{Jr.}~\bibnamefont{A.~A.~Abdumalikov}}, \bibinfo {author}
  {\bibfnamefont{M.}~\bibnamefont{Baur}}, \bibinfo {author}
  {\bibfnamefont{S.}~\bibnamefont{Filipp}}, \bibinfo {author}
  {\bibfnamefont{M.~P.}\ \bibnamefont{da~Silva}}, \bibinfo {author}
  {\bibfnamefont{A.}~\bibnamefont{Blais}},\ and\ \bibinfo {author}
  {\bibfnamefont{A.}~\bibnamefont{Wallraff}},\ }%
  \bibfield{journal}{%
  \bibinfo {journal} {Phys. Rev. Lett.}\ }%
  \textbf{\bibinfo {volume} {106}},\ \bibinfo {pages} {243601} (\bibinfo {year}
  {2011})%
  \bibAnnoteFile{NoStop}{Lang}%
\bibitem{Houck1}%
  \BibitemOpen
  \bibfield{author}{%
  \bibinfo {author} {\bibfnamefont{A.~A.}\ \bibnamefont{Houck}}, \bibinfo
  {author} {\bibfnamefont{D.~I.}\ \bibnamefont{Schuster}}, \bibinfo {author}
  {\bibfnamefont{J.~M.}\ \bibnamefont{Gambetta}}, \bibinfo {author}
  {\bibfnamefont{J.~A.}\ \bibnamefont{Schreier}}, \bibinfo {author}
  {\bibfnamefont{B.~R.}\ \bibnamefont{Johnson}}, \bibinfo {author}
  {\bibfnamefont{J.~M.}\ \bibnamefont{Chow}}, \bibinfo {author}
  {\bibfnamefont{L.}~\bibnamefont{Frunzio}}, \bibinfo {author}
  {\bibfnamefont{J.}~\bibnamefont{Majer}}, \bibinfo {author}
  {\bibfnamefont{M.~H.}\ \bibnamefont{Devoret}}, \bibinfo {author}
  {\bibfnamefont{S.~M.}\ \bibnamefont{Girvin}},\ and\ \bibinfo {author}
  {\bibfnamefont{R.~J.}\ \bibnamefont{Schoelkopf}},\ }%
  \bibfield{journal}{%
  \bibinfo {journal} {Nature}\ }%
  \textbf{\bibinfo {volume} {449}},\ \bibinfo {pages} {328} (\bibinfo {year}
  {2007})%
  \bibAnnoteFile{NoStop}{Houck1}%
\end{thebibliography}

%

\end{document}